\documentclass[11pt]{article}
\usepackage{graphicx}
\usepackage{layout}

\begin{document}

\title{Gravito-electromagnetism versus electromagnetism}
\author{A. Tartaglia, M.L. Ruggiero \\
%EndAName
Dip. Fisica, Politecnico and INFN, Torino, Italy\\
E-mail: angelo.tartaglia@polito.it, matteo.ruggiero@polito.it}
\maketitle

\begin{abstract}
The paper contains a discussion of the properties of the
gravito-magnetic interaction in non stationary conditions. A
direct deduction of the equivalent of Faraday-Henry law is given.
A comparison is made between the gravito-magnetic and the
electro-magnetic induction, and it is shown that there is no
Meissner-like effect for superfluids in the field of massive
spinning bodies. The impossibility of stationary motions in
directions not along the lines of the gravito-magnetic field is
found. Finally the results are discussed in relation with the
behavior of superconductors.
\end{abstract}

\section{Introduction}

While teaching basic physics in universities both for science and
for engineering curricula, the electromagnetic field and the
gravitational field are still treated as completely separated
topics. Even when the essentials of relativity are taught this
separation is kept. Gravitation and electromagnetism are indeed
different: the former is impied in the geometric properties of
space time, the latter is a field living within that geometric
environment on which it reacts back as any other field. It is
however worth evidencing some strong analogies between the two
theories, which inspired many a physicist in the late XIX century.
Even today the similarity between the Maxwell equations on one
side and the linearized Einstein equations on the other, is
intriguing. It would be an error to overlook as well as to
overestimate it. With the appropriate caveats the analogy could
suggest interesting though extremely difficult experiments
exploiting a 'gravitational' Faraday-Henry law. We think that, in
general, students should have a glimpse to the interplay between
classical electromagnetism and General Relativity. A glimpse on
the side of the students means attention and commitment on the
side of the teacher. Here we shall try to summarize and underline
the essentials of the correspondence between the two theories.

Since the early times of general relativity it is known that linearizing the
Einstein field equations in vacuo leads to a form almost identical to the
Maxwell equations of electromagnetism\cite{ruggiero02}. In practice in the
linear approximation the gravitational interaction may be thought of as if
it were the effect of a gravito-electromagnetic field. The gravito-electric
field is the known Newtonian solution. The gravito-magnetic part is an
unexpected contribution much similar to the magnetic field originated by
electric charge currents. There have been many attempts to exploit the
electromagnetic analogy in order to evidence gravito-magnetic effects.
Observational or experimental activities have also been set up to directly
reveal the effects\cite{bct},\cite{iorio02},\cite{ciufolini02a}. What we
would like to discuss here is to what extent the electromagnetic analogy can
be exploited.

Once the analogy has been established there are many consequences one can
draw, for instance a gravito-magnetic induction can be expected from the
analogue of Faraday-Henry law
\begin{equation}
\overrightarrow{\nabla }\wedge \overrightarrow{E}_{g}=-\frac{1}{2}\frac{%
\partial \overrightarrow{B}_{g}}{\partial t}  \label{henry}
\end{equation}
Here $\overrightarrow{E}_{g}$ is the gravito-electric part of the
gravitational field, and $\overrightarrow{B}_{g}$ its gravito-magnetic part.
A time varying gravito-magnetic field will induce a gravito-electric field,
and vice versa.

Of course there are some caveats: $E_{g}$ is a (three)acceleration, $B_{g}$
is the inverse of a time much like an angular velocity; the coupling
parameter is now the mass, and for $\overrightarrow{B}_{g}$ it doubles the
value it has for $\overrightarrow{E}_{g}$ (hence the $1/2$ factor in eq. \ref
{henry}). Furthermore a peculiar difference in sign in the equivalent of the
Amp\`{e}re-Maxwell equation brings about some interesting consequences which
develop in actual inconsistencies when inadvertently exceeding the limits of
the basic approximation. The most important point to be advised of is that
the linearized equations of gravito-electromagnetism are neither really
gauge-invariant nor generally covariant\cite{ruffini}.

Despite the necessary cautions, however, eq. (\ref{henry}) lends a principle
possibility to reveal gravito-magnetism, so it is useful to analyze it
better.

In this paper we shall deduce (\ref{henry}) not from the
linearized Einstein equations, but directly from the equations of
motion of a mass and from the metric tensor. The approximation we
shall use is consequent to the hypothesis that all velocities will
be small as compared with the speed of light, and the field will
be weak enough. We shall then discuss the equivalent of the
Meissner effect for superconductors when a superfluid is taken
into account. As we shall see, the behavior induced in matter by
gravito-magnetism will be different from what happens with
electro-magnetism in superconductors. Finally we shall discuss the
implication that some kinds of stationary motions of a fluid in
gravito-magnetic fields are untenable.

\section{Direct deduction of the gravito-electromagnetic Faraday-Henry law}

Let us consider an axially symmetric stationary space time. Its line element
may be written:
\[
ds^{2}=g_{tt}dt^{2}+g_{xx}dx^{2}+g_{yy}dy^{2}+g_{zz}dz^{2}+2g_{tx}dtdx+2g_{ty}dtdy
\]
If
\begin{eqnarray*}
g_{tx} &=&-B_{g}y/2 \\
g_{ty} &=&B_{g}x/2
\end{eqnarray*}
the space-time contains a constant gravito-magnetic field $B_{g}$ along the $%
z$ axis of a Cartesian coordinate system. Otherwise the space-time is
assumed to be flat, i.e.
\begin{equation}
g_{\mu \nu }=\left(
\begin{array}{llll}
c^{2} & -B_{g}y/2 & B_{g}x/2 & 0 \\
-B_{g}y/2 & -1 & 0 & 0 \\
B_{g}x/2 & 0 & -1 & 0 \\
0 & 0 & 0 & -1
\end{array}
\right)  \label{metrica}
\end{equation}

Strictly speaking, this would be the situation inside a steadily rotating
massive shell. This field has been studied by Lense and Thirring\cite
{lensthir1} soon after the publication of the General Theory of Relativity,
by Brill and Cohen\cite{brill66} in the 1960's, in connection with the
effect of rotating masses on inertial frames, and also, more recently, by
Ciufolini and Ricci\cite{ciufolini02}, who studied the effect of rotation on
the time delay.

It is interesting to compare the metric tensor (\ref{metrica}) with the line
element of Minkowski space time as seen from a steadily rotating reference
frame (rotation about the $z$ axis):
\begin{equation}
ds^{2}=\left( c^{2}-\omega ^{2}x^{2}-\omega ^{2}y^{2}\right)
dt^{2}-dx^{2}-dy^{2}-dz^{2}-2\omega ydxdt+2\omega xdydt  \label{rotante}
\end{equation}
As far as the peripheral rotation speed $\omega \sqrt{x^{2}+y^{2}}$ is
negligible with respect to $c$ (\ref{metrica}) and (\ref{rotante}) coincide
with $B_{g}/2$ playing the role of $\omega $.

Explicitly writing the Christoffels corresponding to (\ref{metrica})
produces:
\[
\begin{array}{ll}
\Gamma _{tx}^{t}=\frac{B_{g}^{2}x}{4c^{2}+B_{g}^{2}\left( x^{2}+y^{2}\right)
} & \Gamma _{ty}^{t}=\frac{B_{g}^{2}y}{4c^{2}+B_{g}^{2}\left(
x^{2}+y^{2}\right) } \\
\Gamma _{tx}^{x}=-\frac{B_{g}^{3}xy}{8c^{2}+2B_{g}^{2}\left(
x^{2}+y^{2}\right) } & \Gamma _{ty}^{x}=\frac{B_{g}(B_{g}^{2}x^{2}+4c^{2})}{%
8c^{2}+2B_{g}^{2}\left( x^{2}+y^{2}\right) } \\
\Gamma _{tx}^{y}=-\frac{B_{g}(B_{g}^{2}y^{2}+4c^{2})}{8c^{2}+2B_{g}^{2}%
\left( x^{2}+y^{2}\right) } & \Gamma _{ty}^{y}=-\frac{B_{g}^{3}xy}{%
8c^{2}+2B_{g}^{2}\left( x^{2}+y^{2}\right) }
\end{array}
\]
Assuming, as said in the Introduction, that all velocities are small when
compared to the speed of light, we can write, at the lowest approximation
order, $ds=cdt$. Furthermore, under the same assumption the analogy between $%
B_{g}$ and an angular speed suggests that $B_{g}\sqrt{\left(
x^{2}+y^{2}\right) }<<c$. The significant Christoffels are then reduced to
\[
\begin{array}{l}
\Gamma _{ty}^{x}=\frac{B_{g}}{2} \\
\Gamma _{tx}^{y}=-\frac{B_{g}}{2}
\end{array}
\]
The components of the covariant four-acceleration $\textsl{a}$
(same approximation as above) are:
\begin{eqnarray}
&&\textsl{a}^{0}\simeq 0  \nonumber \\
\textsl{a}^{x}
&=&\frac{d^{2}x}{c^{2}dt^{2}}+B_{g}\frac{dy}{c^{2}dt}
\nonumber \\
\textsl{a}^{y}
&=&\frac{d^{2}y}{c^{2}dt^{2}}-B_{g}\frac{dx}{c^{2}dt}
\label{covaccel} \\
\textsl{a}^{z} &=&\frac{d^{2}z}{c^{2}dt^{2}}  \nonumber
\end{eqnarray}

Suppose now that a material point is constrained to move so that:
\begin{eqnarray}
x &=&R\cos \theta  \nonumber \\
y &=&R\sin \theta \cos \left( \Omega t\right)  \label{motion} \\
z &=&R\sin \theta \sin \left( \Omega t\right)  \nonumber
\end{eqnarray}

In practice we are assuming that, whatever the probe we are using is, it is
free to move along a rigidly steadily rotating circular ring; otherwise
stated, we are considering a fluid inside a rotating anular tube. $R$ is the
radius of the ring, $\Omega $ is its angular velocity; rotation takes place
about the $x$ axis; $\theta \left( t\right) $ is an angular parameter
showing the position along the ring. This is the same configuration as for a
typical electromagnetic induction experiment in an alternator.

Let us now introduce (\ref{motion}) into (\ref{covaccel}). Since the motion
is not geodesic the result will in general be different from zero; it will
correspond to the components of the four-force per unit mass constraining
the probe to stay in the ring.

Explicitly, and considering the only space components:

It will be
\begin{eqnarray*}
c^{2}\textsl{a}^{x} &=&{a}^{x}=-R\frac{d^{2}\theta }{dt^{2}}\sin
\theta
-R\left( \frac{d\theta }{dt}\right) ^{2}\cos \theta +B_{g}R\left( \frac{%
d\theta }{dt}\cos \theta \cos \left( \Omega t\right) -\Omega \sin \theta
\sin \left( \Omega t\right) \right) \\
c^{2}\textsl{a}^{y} &=&{a}^{y}=R\frac{d^{2}\theta }{dt^{2}}\cos
\theta \cos \left( \Omega t\right) -R\left( \frac{d\theta
}{dt}\right) ^{2}\sin \theta \cos \left( \Omega t\right) -2R\Omega
\frac{d\theta }{dt}\cos \theta \sin
\left( \Omega t\right) \\
&&-R\Omega ^{2}\sin \theta \cos \left( \Omega t\right) +B_{g}R\frac{d\theta
}{dt}\sin \theta \\
c^{2}\textsl{a}^{z} &=&{a}^{z}=R\frac{d^{2}\theta }{dt^{2}}\cos
\theta \sin \left( \Omega t\right) -R\left( \frac{d\theta
}{dt}\right) ^{2}\sin \theta \sin \left( \Omega t\right) +2R\Omega
\frac{d\theta }{dt}\cos \theta \cos
\left( \Omega t\right) \\
&&-R\Omega ^{2}\sin \theta \sin \left( \Omega t\right)  \label{accelapprox}
\end{eqnarray*}
The constraints represented by the walls of the circular ring of course
react to the forces orthogonal to the wall. The only unconstrained direction
will be the one tangent to the ring. For that direction we can write:
\begin{equation}
{a}^{x}\sin \theta -{a}^{y}\cos \theta \cos \left( \Omega t\right) -{a}%
^{z}\cos \theta \sin \left( \Omega t\right) =0  \label{tangent}
\end{equation}

Finally, introducing (\ref{accelapprox}) in (\ref{tangent}), the
approximated expression for the angular acceleration along a rotating ring
in presence of a gravito-magnetic field will be:

\[
\frac{d^{2}\theta }{dt^{2}}\simeq -B_{g}\Omega \sin \left( \Omega t\right)
\sin ^{2}\theta +\Omega ^{2}\sin \theta \cos \theta
\]
The second term corresponds to the transverse contribution of the
centrifugal acceleration. The other one is the true gravito-magnetic
induction contribution, if we like to call it so.

The linear acceleration will of course be
\[
R\frac{d^{2}\theta }{dt^{2}}\simeq -B_{g}R\Omega \sin \left( \Omega t\right)
\sin ^{2}\theta +R\Omega ^{2}\sin \theta \cos \theta
\]
Integrating along the length of the ring one has a work per unit mass
(gravito-electromotive force)
\begin{eqnarray}
\mathcal{F} &=&\int_{0}^{2\pi }R^{2}\left( -B_{g}\Omega \sin \left( \Omega
t\right) \sin ^{2}\theta +\Omega ^{2}\sin \theta \cos \theta \right) d\theta
\label{finale} \\
&=&-\allowbreak \pi R^{2}\Omega B_{g}\sin \Omega t  \nonumber
\end{eqnarray}
This in practice is the equivalent of Faraday-Henry law for classical
electrodynamics, in integral form; the usual application of Stoke's theorem
brings about the differential form (\ref{henry}).

In fact if we consider the $g_{ti}$'s of our metric tensor as the components
of a gravito-magnetic vector potential, the corresponding gravito-magnetic
field will be $B_{g}$ and (\ref{finale}) is the time derivative of the flux
of $\overrightarrow{B}_{g}$ across the area of the ring. Furthermore, we
have also practically checked the validity of a Lorentz-like force law $%
\overrightarrow{{a}}=\overrightarrow{v}\wedge \overrightarrow{B}_{g}$.
Indeed, considering (\ref{covaccel}) and recalling that the costrained
motion has only $v_{y}$ and $v_{z}$ components of the velocity, and the
gravito-magnetic field is along the $z$ axis we see that the $B_{g}$
dependent acceleration has the form of the typical vector product in the
Lorentz force formula.

Up to this moment no troubles arise, provided the approximation conditions
are satisfied.

If one considers the ponderomotive force corresponding to (\ref{finale}) we
can think of a principle means to reveal the existence of a gravito-magnetic
field. This could happen in a superconductor ring equipped with a Josephson
junction, though in that case there would be a serious problem with the pure
magnetic effects. Non inertial effects in superconductors, according to the
scheme we mentioned, have been studied by Fisher et al. \cite{fisher}.
Another possibility would be to consider a superfluid flow in a ring shaped
tube; in this case one would avoid troubles with magnetic interactions.

\section{Gravitational Meissner effect}

Considering superconducting devices implies considering the typical Meissner
effect too. Is there a gravito-magnetic analog of the Meissner effect? The
answer we can find in the literature is partly yes, as in DeWitt\cite{dewitt}%
, Li-Torr\cite{li},\cite{li1}\footnote{%
However, the papers by Li-Torr have been criticized by Kowitt\cite{kowitt},
who pointed out that they \textit{\ grossly overestimated} the magnitude of
the effects; we may add that the paper \cite{li} contains a little mistake:
the typical $1/2$ factor expressing the difference between the
gravito-electric and gravito-magnetic mass is missing.}; there however a
superconductor was analyzed so that there was an interplay between
electromagnetic and gravito-electromagnetic interactions. The conclusion was
that the gravito-magnetic, as well as the magnetic field, weaken when
penetrating into the bulk of a superconductor, though the typical
penetration length for the gravito-magnetic part is so big that the effect
is practically irrelevant. If however one wants to consider a real analog of
the pure Meissner effect, one should treat the pure gravito-magnetic case.
To be more definite, we should envisage a situation where matter can flow
without friction in response to a gravito-electromagnetic field, i.e. matter
in a pure superfluid state.

Let us start with the Maxwell-like equations for the gravito-electromagnetic
field:
\begin{equation}
\left\{
\begin{array}{c}
\overrightarrow{\nabla }\cdot \overrightarrow{E}_{g}=-4\pi G\rho \\
\overrightarrow{\nabla }\cdot \overrightarrow{B}_{g}=0 \\
\overrightarrow{\nabla }\wedge \overrightarrow{E}_{g}=-\frac{1}{2}\frac{%
\partial \overrightarrow{B}_{g}}{\partial t} \\
\overrightarrow{\nabla }\wedge \overrightarrow{B}_{g}=-\frac{8\pi G}{c^{2}}%
\overrightarrow{j}_{g}+\frac{2}{c^{2}}\frac{\partial \overrightarrow{E}_{g}}{%
\partial t}
\end{array}
\right.  \label{pseudomax}
\end{equation}
So far we have seen that the third equation is O.K. deducing it, in its
integral form, directly from the equations of motion. As for the other ones,
they are obtained from the linearized Einstein equations\cite{mashhoon99}.
Let us concentrate on the last pair of equations and try and solve them in
general.

In this case we must also consider that $\overrightarrow{E}_{g}$ is the
three-acceleration effective in changing the absolute value of the
three-velocity of matter along a loop, the one entering the very definition
of the matter current density. In weak field conditions, it is
\begin{equation}
\overrightarrow{j}_{g}=\frac{\rho \overrightarrow{v}}{\sqrt{1-\frac{v^{2}}{%
c^{2}}}}  \label{gei}
\end{equation}
As far as the velocity of the matter flow is small enough and we can assume
that the self-perturbation of the matter density $\rho $ is negligible, we
have
\[
\frac{\partial \overrightarrow{j}_g}{\partial t}=\rho \overrightarrow{E}_{g}
\]
Let us now differentiate the last Maxwell-like equation with respect to
time, then introduce into it the third one:
\begin{equation}
\overrightarrow{\nabla }\wedge \overrightarrow{\nabla }\wedge
\overrightarrow{E}_{g}=\frac{4\pi G}{c^{2}}\rho \overrightarrow{E}_{g}-\frac{%
1}{c^{2}}\frac{\partial ^{2}\overrightarrow{E}_{g}}{\partial t^{2}}
\label{eq-in-E}
\end{equation}
This corresponds to
\[
-4\pi G\overrightarrow{\nabla }\left( \rho \right) -\nabla ^{2}%
\overrightarrow{E}_{g}=\frac{4\pi G}{c^{2}}\rho \overrightarrow{E}_{g}-\frac{%
1}{c^{2}}\frac{\partial ^{2}\overrightarrow{E}_{g}}{\partial t^{2}}
\]
Suppose that inside matter it is $\overrightarrow{\nabla }\left( \rho
\right) =0$; the equation reads
\[
\nabla ^{2}\overrightarrow{E}_{g}=\frac{1}{c^{2}}\frac{\partial ^{2}%
\overrightarrow{E}_{g}}{\partial t^{2}}-\frac{4\pi G}{c^{2}}\rho
\overrightarrow{E}_{g}
\]
Introducing the time Fourier transform $\overrightarrow{\mathcal{E}}\left(
r,\omega \right) =\int_{-\infty }^{\infty }\overrightarrow{E}_{g}e^{i\omega
t}dt$ we have
\[
\nabla ^{2}\overrightarrow{\mathcal{E}}=-\frac{\omega ^{2}}{c^{2}}%
\overrightarrow{\mathcal{E}}-\frac{4\pi G}{c^{2}}\rho \overrightarrow{%
\mathcal{E}}
\]

The solution is
\begin{eqnarray*}
\overrightarrow{\mathcal{E}} &=&\overrightarrow{\mathcal{E}}_{0}\left(
\omega \right) e^{ikr} \\
k &=&\frac{1}{c}\sqrt{\omega ^{2}+4\pi G\rho }
\end{eqnarray*}
Here $r$ stands for a space coordinate orthogonal to the surface of the
material. Introducing this result into the third equation of (\ref{pseudomax}%
) we see of course that the space distribution of the gravito-magnetic $%
\overrightarrow{B}_{g}$ field is of the same type as for the
gravito-electric field. It oscillates in space, rather then being damped.
There is no equivalent of the electromagnetic plasma frequency:\ the
exponential $e^{ikr}$ never becomes real. The wavelength of the space
oscillation in static conditions ($\omega =0$) is
\[
\lambda =c\sqrt{\frac{\pi }{G\rho }}
\]

In practice the numeric value of $\lambda $ for ordinary situations would be
$\sim 10^{12}$ m. The difference with respect to a damped trend is
substantially irrelevant, however we can state that in principle in
superfluids there is no analog of the Meissner effect in superconductors.
Differently phrased, recalling a remark by Pascual-Sanchez \cite{pascual},
superfluids in gravito-magnetic fields display a paramagnatic-like behavior
rather than a diamagnetic-like one.

\section{Inconsistencies of the gravito-electromagnetic analogy}

The result regarding the Meissner effect may induce the suspect that
something is wrong with the last Maxwell-like equation. In fact the behavior
we have seen depends on the '$-$' sign in front of the matter current
density in the last equation of (\ref{pseudomax}). That little '$-$'
produces indeed other apparently anomalous, not to say incongruous,
consequences. Essentially we can say that the lines of force of $%
\overrightarrow{B}_{g}$ circulate around the matter flux in clockwise sense,
opposite to what happens with a magnetic field with respect to an electric
current.

Let us begin with a simple example. The gravito-magnetic Faraday-Henry law (%
\ref{finale}) is formally equivalent to the induction law for the
electromagnetic fields. So, we can reasonably expect that a gravito-magnetic
time-varying flux $\Phi _{g}$ induces a gravito-electric field $%
\overrightarrow{E}_{g}$. We born ourselves to the simplest situation one can
imagine.

\begin{figure}[tp]
\begin{center}
\includegraphics[width=8cm,height=8cm]{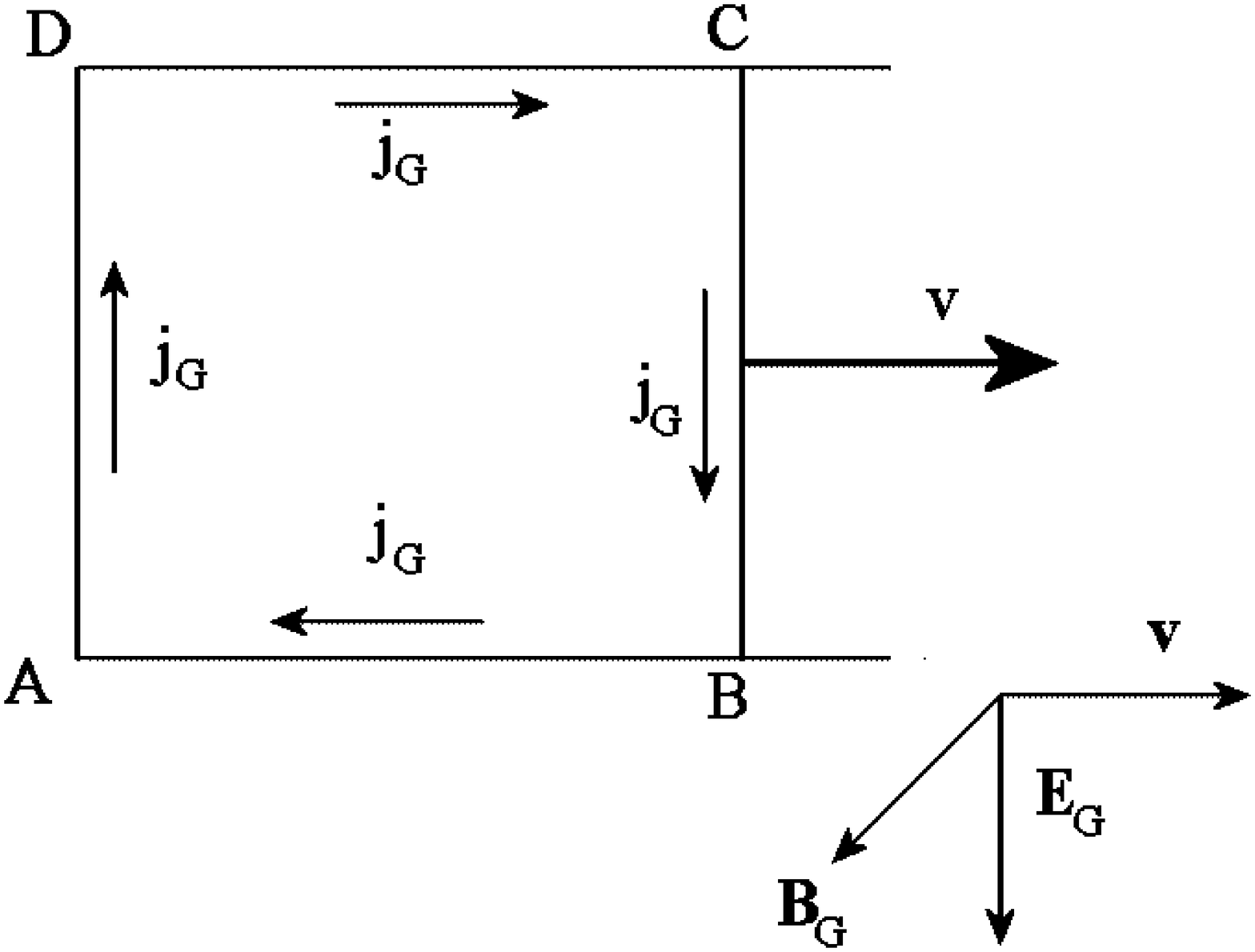}
\end{center}
\caption{Scheme of a fluid circuit with a moving arm, immersed in a
gravitomagnetic field oriented orthogonally with respect to the figure,
toward the observer.}
\label{henry1}
\end{figure}

%\FRAME{ftbpFU}{4.6673in}{3.5045in}{0pt}{\Qcb{Scheme of a fluid circuit with
%a moving arm, immersed in a gravitomagnetic field oriented orthogonally
%with respect to the figure, toward the observer.}}{\Qlb{henry1}}{figura.eps}{%
%\special{language "Scientific Word";type "GRAPHIC";maintain-aspect-ratio
%TRUE;display "USEDEF";valid_file "F";width 4.6673in;height 3.5045in;depth
%0pt;original-width 11.4004in;original-height 8.5416in;cropleft "0";croptop
%"1";cropright "1";cropbottom "0";filename 'figura.eps';file-properties
%"XNPEU";}}

Suppose that a massive fluid is constrained to move in a circuit $ABCD$,
where the $CB$ side is moving with constant velocity $V$; everything is
immersed in a gravito-magnetic field $\overrightarrow{B}_{G}$, perpendicular
to the plane and directed toward the reader (see figure \ref{henry1}). The
massive particles in the segment $CB$ are acted upon by a Lorentz-like force
\[
\overrightarrow{F}=m\overrightarrow{V}\wedge \overrightarrow{B}_{g}
\]
so they start moving under the influence of the induced gravito-electric
field $\overrightarrow{E}_{g}=\frac{\overrightarrow{F}}{m}$ and a mass
current density $\overrightarrow{j}_{g}$ appears, as depicted.

According to eq. $\overrightarrow{\nabla }\wedge \overrightarrow{B}_{g}=-%
\frac{8\pi G}{c^{2}}\overrightarrow{j}_{g}+\frac{2}{c^{2}}\frac{\partial
\overrightarrow{E}_{g}}{\partial t}$ this current is the source for a
gravito-magnetic field $\overrightarrow{b}_{g}$, which satisfies the
gravito-magnetic Ampere law
\[
\oint \overrightarrow{b}_{g}\cdot \overrightarrow{dl}=-\frac{8\pi G}{c^{2}}%
i_{g}
\]
Of course $i_{g}$ is the total mass current in the circuit.

Because of the "minus" sign, this field is directed in the same direction as
the initial $\overrightarrow{B}_{g}$ field, whose varying flux induces the
current! In practice the system would diverge. Of course an increasing
current leads soon to a violation of the linearization conditions, but the
relevant fact is that, in a sense, the approximation is unstable.

\section{Discussion}

The fact that the linearization of the Einstein equations, which produces
the Mawell-like equations (\ref{pseudomax}) has strong limitations is well
known\cite{MTW},\cite{ruffini}, since it leads to a non self consistent
theory. The point which has been considered in the literature concerns the
issue of energy, since gravitational forces \textit{"do no significant work"}
and the energy-stress tensor $T^{\mu\nu}$ is conserved independently of the
action of the gravitational fields.

Gravitational energy, however, is in general an open problem even in the
exact theory.

Here we can see that the energy balance, in the style of special relativity,
is indeed satisfied.

We have shown, with the simple example represented in figure \ref{henry1},
that the gravito-electromagnetic induction would produce an indefinitely
increasing matter flux in a fluid. This result is of course paradoxical in a
non-relativistic approach. However, as far as the speed of matter particles
(the flux, indeed) increases, purely special relativistic effects cannot be
neglected. In practice to keep the translation speed constant requires a
force. Using special relativistic formulas we expect the force (in the
direction of $\overrightarrow{V}$) to be
\begin{eqnarray}
F &=&m\frac{d}{dt}\left( \frac{V}{\sqrt{1-\frac{V^{2}+v^{2}}{c^{2}}}}\right)
\nonumber \\
&=&\frac{mV}{\left( 1-\frac{V^{2}+v^{2}}{c^{2}}\right) ^{3/2}}\frac{v}{c^{2}}%
\frac{dv}{dt}  \label{force} \\
&=&\frac{mV}{1-\frac{V^{2}+v^{2}}{c^{2}}}\frac{j}{\rho c^{2}}E  \nonumber
\end{eqnarray}

In the formula, $m$ is the rest mass of the flowing matter in the moving arm
of the circuit; the mass of the container (that is also moving) has been
neglected, because, under the hypothesis $V=$ constant its time derivative
would be zero. The small $v$ stands for the velocity of the flowing matter,
which, in our example, is orthogonal to the translational motion of the $CB$
arm. $E$ is the acceleration induced in the flow by the gravito-magnetic
interaction.

A consequence of what we have seen is that no static fluid flow orthogonal
to $\overrightarrow{B}_{g}$ is possible since it would require an
asymptotically infinite force and would absorb an infinite energy. Nothing
can be kept in motion in a gravito-magnetic field if not along the lines of
the field.

The same conclusion can be attained in the case of a closed fluid circuit
set to rotation about an axis not aligned with the field. In other words a
fluid gyroscope cannot maintain its angular momentum constant if not aligned
with the field.

In drawing this conclusion we have released the condition of small
velocities but not the one of having a weak external field. Of course if the
system actually reaches relativistic conditions we can no longer neglect the
back reaction on the structure of the global field, however we do not expect
the non-linearities to modify qualitatively the result concerning the
untenability of stationary motions in a gravito-magnetic field.

Actually all we have said is referred to uncharged flowing matter. The
situation changes when electric currents or supercurrents are considered. As
already said, Li and Torr\cite{li} showed that in a superconductor magnetic
and gravito-magnetic fields are indeed coupled and the final result is that,
in the material, both fields decay exponentially with increasing depth.

In our case we can describe the situation as follows. In a pure
superconductor a varying magnetic field $\overrightarrow{B}$ induces a
supercurrent whose back reaction is a field that entirely compensates the
changes in $\overrightarrow{B}$. If a varying external gravito-magnetic
field is also present, its effect is to contribute to a mass flow in the
superconductor whose effect would lead to a continuous increase of the flow
itself. However, since the matter flow is also an electric current, the
magnetic field of the current will be strong enough to stabilize the system.
Summing up, a superconductor loop in a varying gravito-magnetic field will
indeed reach a state of dynamic equilibrium with a supercurrent (then the
corresponding magnetic field) a bit stronger than it would be the case
without the gravito-magnetic interaction. Unfortunately, if the
gravito-magnetic field is the one of the Earth its effect is extremely small
\cite{tajmar}.

We think that our description of the gravito-magnetic induction,
compared and contrasted with the electro-magnetic one, can help in
shedding light on the properties of the gravitational field of
rotating bodies, to the benefit of graduate students who are
interested in the subject, but seldom have the opportunity to
adequately approach it. Even for undergraduate students, the
comparison of electro-magnetism and gravitation, if presented
together with a historical sketch of the evolution of both
theories, should lead to a better understanding of the phenomena
they describe.

\end{document}